\documentclass[12pt]{iopart}
\usepackage{epsfig}
\usepackage{graphicx}
\usepackage{iopams}
\usepackage{hyperref}
\usepackage{amssymb}
\usepackage{bm,bbm}
\usepackage{color}
\usepackage{threeparttable}%
\hypersetup{colorlinks,citecolor=blue,filecolor=blue,linkcolor=blue,urlcolor=blue}
\begin{document}
\title{Topological end states in a one-dimensional spatially modulated interaction spinless fermion model}
\author{Zheng-Wei Zuo$^{1,2}$, Da-wei Kang$^1$,  and Liben Li$^1$}
\address{$^1$School of Physics and Engineering, and Henan Key Laboratory of Photoelectric Energy Storage Materials and Applications, Henan University of Science and Technology, Luoyang 471023, China}
\address{$^2$Department of Physics, The Pennsylvania State University, University Park, Pennsylvania 16802, USA}

\begin{abstract}
The effect of spatially modulated interaction on quantum phase transition in one-dimensional interacting spinless fermion system is theoretically investigated by exact diagonalization and density matrix renormalization group method. Our calculations show that the periodically modulated interaction can drive the spinless fermion system into topological charge density wave state. The topological state is encoded by quasiparticle end states and the fractional quantized $e/2$ end charges, and characterized by Berry phase and Chern number. The quasiparticle energy spectra as a function of modulated interaction period appears a stunning fractal-like structure. For the quasi-periodic case, the topological phase transition can also occur. In a word, the spatially modulated interaction can be a new elegant avenue towards realizing to interacting topological phases.

\end{abstract}


\hspace{4pc}\noindent{Keywords: Topological phases, Density matrix renormalization group}

\hspace{9pc}\noindent{Topological Mott insulator, Exact diagonalization}

%\hspace{9pc}\noindent{Renormalization group}

\maketitle

\section{Introduction}

Topological phases of matter have been one of the central topics of the condensed matter physics~\cite{Hasan10RMP,QiXL11RMP,Franz15RMP}. Among a number of such topological quantum systems, topological states in gapped free-fermion systems are the well-understood examples. As we know, the fermions and bosons interaction induces intriguing and exotic phases\cite{WangQR18PRX,WangC14SCI,PotterAC16PRX,DominicVE16PRB,FidkowskiL10PRB,FidkowskiL11PRB,TurnerAM11PRB,ChenXie11PRB1,ChenXie11PRB2}, which can not be realized in free-fermion and free-boson systems. The interacting boson and fermion systems have become a fascinating theoretical direction in recent years\cite{Stern10NT,MaciejkoJ15NTP,SternA16ARCMP,AliceaJ16ARCMP,SenthilT15ARCMP,HanssonTH17RMP,WenXG17RMP}. What is more, because of its simplicity, the one-dimensional (1D) interacting topological quantum systems have been the subject of many studies\cite{FidkowskiL10PRB,FidkowskiL11PRB,TurnerAM11PRB,ChenXie11PRB1,ChenXie11PRB2,XuZH13PRL,XuZH13PRB,GuoHM15PRB,ZhuSL13PRL,GrusdtF13PRL,DengXL14PRA,LiTH15PRB,MatsudaF14JPSJ,ZengTS16PRB,HuLH19arXiv}. For example, the electron interaction can radically modify the topological classification of the 1D free fermion systems\cite{FidkowskiL10PRB}. The ID fermion-Hubbard\cite{XuZH13PRL,XuZH13PRB,GuoHM15PRB}, bose-Hubbard models\cite{ZhuSL13PRL,DengXL14PRA,LiTH15PRB,ZengTS16PRB,GrusdtF13PRL,MatsudaF14JPSJ}, Luttinger liquid systems\cite{RuhmanJ15PRL,Keselman15PRB,KainarisN15PRB,MontorsiA17PRB,RuhmanJ17PRL}, interacting Floquet systems\cite{PotterAC16PRX,KeyserlingkCW16PRB,KeyserlingkCW16PRB2,DominicVE16PRB}, and Schwinger model\cite{MagnificoG19PRD,MagnificoG19PRB} are the prototypical systems. The analytical and numerical methods such as bosonization technique\cite{GogolinAO98Book,GiamarchiT04Book}, group cohomology\cite{ChenXie11PRB1}, exact diagonalization\cite{ZhangJM10EJP,RaventosD17JPB,Weinberg17SPP}, and density matrix renormalization group (DMRG) method\cite{White92PRL,Schollwock05RMP,Schollwock11AP} have been used widely for studying these topological strongly correlated matters.

Symmetry plays a key role in topological phases. The  $K$-theory and group theory have been used to classify topological insulators, semimetal, and superconductors under the internal (nonspatial) symmetry and space group~\cite{Kitaev09AIP,SchnyderAP08PRB,Slager12NTP,KruthoffJ17PRX,PoHC17NTC,BradlynB17NT}. On the other hand, the superlattice potential\cite{LangLJ12PRL,ZhuSL13PRL,XuZH13PRL,XuZH13PRB,LiTH15PRB,DengXL14PRA,ZengTS16PRB,ParkJH16PRB,ThakurathiM18PRB,ZuoZW18PRB}, periodically modulated hopping\cite{SuWP79PRL,GaneshanS13PRL,GrusdtF13PRL,MatsudaF14JPSJ,GuoHM15PRB}, and periodically driven field\cite{OkaT09PRB,Kitagawa10PRB,LindnerNH11NTP,PotterAC16PRX,KeyserlingkCW16PRB,KeyserlingkCW16PRB2,DominicVE16PRB} have been the simple and elegant avenues towards realizing fascinating topological phases. The Su-Schrieffer-Heeger (SSH) model\cite{SuWP79PRL}, diagonal and off-diagonal Aubry-Andr\'e-Harper (AAH) model\cite{LangLJ12PRL,GaneshanS13PRL}, and topological Floquet systems \cite{OkaT09PRB,Kitagawa10PRB,LindnerNH11NTP} with their counterpart interaction systems are object of considerable theoretical and experimental interest. Most work of these topological correlated systems focus on the homogeneous interaction terms. It is of fundamental interest to investigate the topological properties for the inhomogeneous/anisotropic interaction cases. In this paper, we consider the effect of spatially modulated interaction. It is natural to ask if analogous behaviors and topological phase transitions can occur in spatially modulated interaction systems. Is any new and emerging physics occurring such as nontrivial topological features? If so, how to identify and characterize the phenomenon? Currently, there is a few study on the spatially modulated interaction systems\cite{PaivaT96PRL,PaivaT98PRB,KogaA13JPSJ,MendesST13PRB}, which mainly discuss the Mott insulating (spin/charge density wave states) properties of one-dimensional Hubbard superlattice. Here, by way of exact diagonalization, DMRG algorithm calculations, and bosonization technology, we take a one-dimensional spatially modulated interaction spinless fermion system as an example and try to answer these questions. We firstly demonstrate that the periodically modulated interaction can induce the topological phase transitions. There are fractional quantized $\pm e/2$ end charges when the system is in topological Mott (charge density wave) states. The quasiparticle energy spectra as a function of modulated interaction period parameter appears fractal-like structure. When the dimer interaction appears, the system is similar to the single-particle SSH model under mean-field approximation. For the quasi-periodically modulated interaction, the topological Mott states still exist.

The rest of the paper is organized as follows. In Sec. \ref{Hamiltonian}, we provide a brief introduction to the Hamiltonian of system. Then, we analytically and numerically demonstrate that the dimer-interaction can drive the system into the topological charge density wave state according to the bosonization technology, exact diagonalization, and DMRG algorithm calculations. In Sec.\ref{PhaseButterfly}, we systematically analyze the topological phase transition and fractal-like structure of quasiparticle energy spectra for different periodically modulated interaction period parameter. In Sec. \ref{Quasiperiod}, we investigate the effect of quasi-periodically modulated interaction on the topological properties for the system. Section \ref{Conclusion} is devoted to conclusions and an outlook.

\section{Model and Result}

\subsection{Topological charge density wave state}\label{Hamiltonian}

Our system can be described by the spinless fermion on a one-dimensional spatially modulated interaction lattice of linear size $La$ ($a=1$ being the lattice spacing)
\begin{equation}
H=\sum_{j}^{L-1}\left[  -t\left(  c_{j}^{\dagger}c_{j+1}+h.c.\right)+V_{j}\hat{n}_{j}\hat{n}_{j+1}\right]  \label{AAHModel}
\end{equation}
where $c_{j}^{\dagger}$ ($c_{j}$) is the creation (annihilation) operator of spinless fermion on site $j$, the operator $\hat{n}_{j}=c_{j}^{\dagger}c_{j}-1/2$. $t$ represents the nearest neighbor hopping strength, where we set to unit hereafter. $V_{j}=V\left[  1+\lambda\cos\left(  2\pi\alpha j+\delta\right) \right]  $ stands for the spatially modulated nearest-neighborhood electron-electron interaction strength with overall amplitude $V$. Here $\lambda$ is the modulation amplitude, and $\delta$ is an arbitrary phase parameter. The modulation period $\alpha=p/q$ ($p, q$ are mutually prime integers for $\alpha$ rational number). As we know, the system is gapped and gapless for $\left\vert V\right\vert >2$ and $\left\vert V\right\vert <2$, respectively when the modulation amplitude $\lambda=0$\cite{GiamarchiT04Book}.

In the following, the filling factor is defined as $\nu=N/L$ with $N$ the number of spinless fermions. Before tackling the general case, we firstly focus on the detailed description of the $\delta=0$ and $\nu=\alpha=1/2$, meaning at the half-filling case with $k_{F}=\pi/2$. We shall firstly approach the Hamiltonian by bosonization technology\cite{GiamarchiT04Book}. In the continuum limit, the effective low-energy Hamiltonian in term of boson filed is written as follows:
\begin{eqnarray}
H = & \frac{1}{2\pi}\int dx\left[  \left(  uK\right)  \left(  \pi\Pi\left(x\right)  ^{2}\right) +\frac{u}{K}\left(  \nabla\phi\left(  x\right)  \right)^{2} \right]  -\nonumber\\
&  \frac{1}{2\pi\alpha_{0}}\int dx\left[ \Lambda\cos\left(  2\phi(x)\right)+\frac{V}{ \pi\alpha_{0}}\cos\left( 4\phi(x)\right) \right]
\label{bosonization}
\end{eqnarray}
where $\phi(x)$ is boson field related to the long wavelength part for the fermion density $\rho=-\nabla\phi\left(  x\right)  /\pi$, $\Pi\left(x\right)$ is the momentum conjugate to this boson field. The $u$ is the renormalized Fermion velocity and The $K$ controls the behaviors of the various correlation functions. The $\Lambda=V\lambda/4$ and $\alpha_{0}$ is a short distance cutoff of the order of the lattice spacing.

The renormalization group equations for various coupling terms of the Hamiltonian read\cite{ZangJ95PRB,Orignac04EPJB,KadanoffLP80PRB}
\begin{eqnarray}
\frac{dy_{\Lambda}}{d\ln b} & =y_{\Lambda}\left(  2-K+y_{V}\right)
\label{Lamda-RG}\\
\frac{dy_{V}}{d\ln b}  &  =y_{V}\left(  2-4K\right)  +\frac{y_{\Lambda}^{2}
}{2}\label{V-RG}\\
\frac{dK}{d\ln b}  &  =-K^{2}\left(  y_{\Lambda}^{2}+y_{V}^{2}\right)
\label{K-RG}\
\end{eqnarray}
where $y_{\Lambda}= \Lambda/ \pi u $, $y_{V}=V/ \pi u $, and $b$ is a scale factor. The scaling diagram can be calculated numerically by ordinary differential equations integrations. Here we only discuss the main results. First, we choose the $V=1$, and $0 \leq \lambda<1 $. According to these scaling equations, calculations show that the scaling process is dominated by $y_{\Lambda}$. Thus, the dimer-interaction is the dominant process and induces an energy gap, which opens in the excitation spectrum of the $\phi$ field and leads to an insulating charge density wave (CDW) state for the spinless fermion system.

\begin{figure}[ptbh]
\center{\includegraphics[scale=0.6]{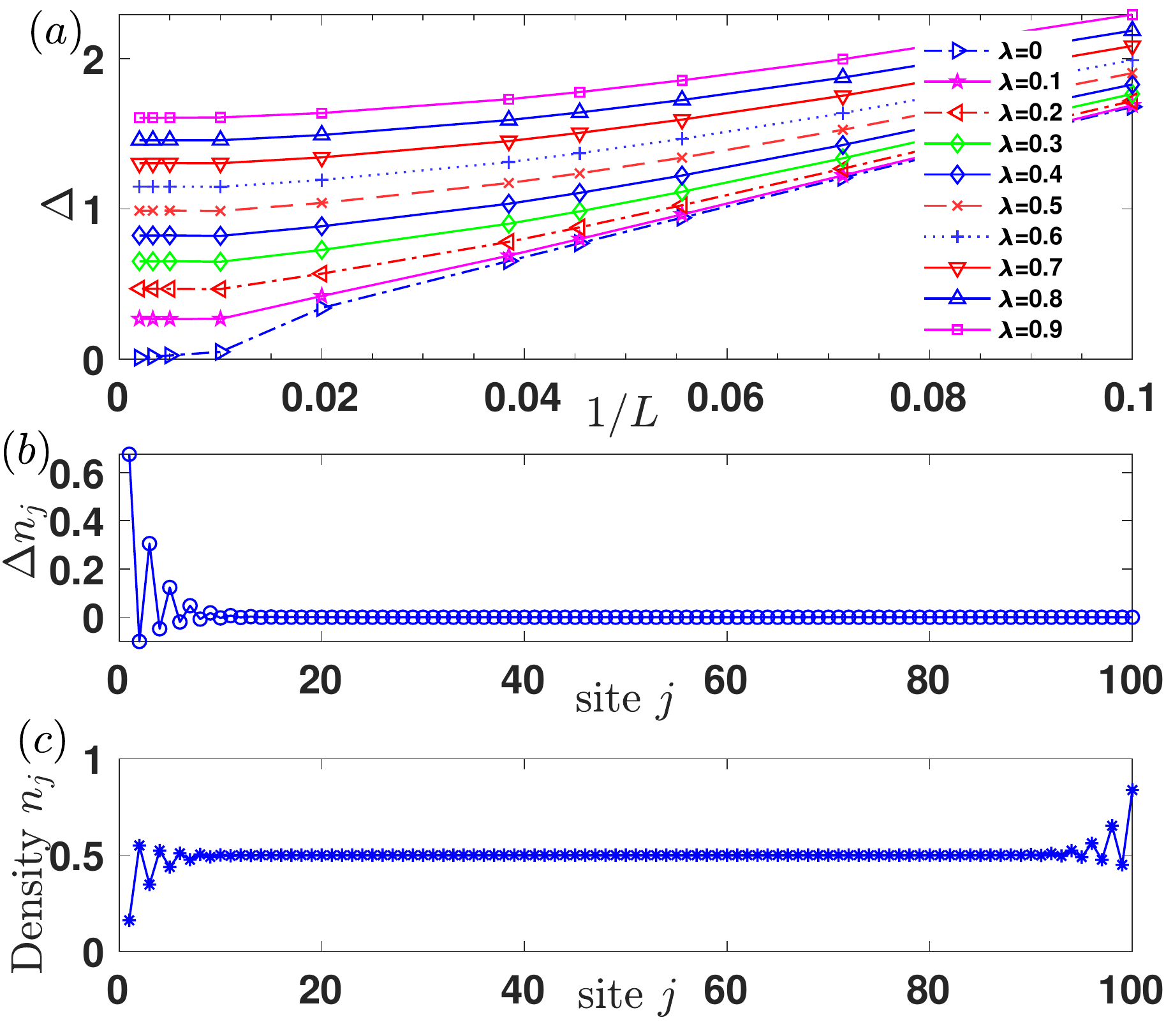}}
\caption{(Color online)$(a)$The finite-size analysis of single-particle charge gap with different $\lambda$, and $V=1, \nu=1/2, \delta=0$; $(b-c)$ The density distribution of the quasiparticle, and electron density for the system with $L=100, V=1, \delta=0, \lambda=0.9, \nu=1/2,N=50$. }
\label{Fig1}
\end{figure}

To verify the insulator state, we calculate the system size dependence of single-particle charge gap under period boundary condition (PBC) using the DMRG method based on the ITensor library\cite{ITensor}.  The single-particle charge gap is defined as $\Delta_L^N=E_0(L, N+1)+E_0(L, N-1)-2E_0(L, N)$, where $E_0(L, N)$ denotes the the ground-state energy with $L$ lattice sites, $N$ particles number. In thermodynamic limit, it will normally be zero in gapless state, while become finite in insulating states. The evolution of the charge gap versus the inverse of the lattice size with various modulation amplitudes $\lambda$ from 0 to 0.9 is plotted in Fig.\ref{Fig1}$(a)$.  From the system size dependence of the charge gap, we can see that the charge gaps are zero for $\lambda=0$ and finite for nonzero $\lambda$ in the thermodynamic limit. So, the system evolves from gapless state (metal state) to the gapped state (CDW state) and the quantum phase transition takes place when the dimer-interaction appears. Thus, our DMRG result is consistent with that of the bosonization technology. We numerically and analytically confirm that the system went through a quantum phase transition from a gapless state to the gap state when the periodically modulated (dimer) interaction turns on. From these scaling equations \ref{Lamda-RG}-\ref{K-RG}, we can see that this quantum phase transition from gapless phase to gap phase is of Kosterlitz-Thouless type.

Does the insulating state have intriguing properties? Next, we show that this Mott (CDW) gap state is topological. In the first place ,we introduce the density distribution of the quasiparticle
\begin{equation}
\Delta n_{j}(N)=\left\langle \psi_{N+1}^{g}\left\vert n_{j}\right\vert \psi_{N+1}^{g}\right\rangle -\left\langle \psi_{N}^{g}\left\vert n_{j}\right\vert\psi_{N}^{g}\right\rangle \label{distributionQP}
\end{equation}
where $\psi_{N}^{g}$ denotes the ground-state wave function of the system with $N$ fermion atoms for open boundary condition (OBC), $n_{j}= c_{j}^{\dagger}c_{j}$. In Fig.\ref{Fig1}$b$, we can see the quasiparticle is mainly localized at one end of chain when lattice size $L=100$, modulation amplitude $\lambda=0.9$, particles number $N=50$.  Similar quasiparticle density distribution is obtained with alternating nearest-neighbor interaction strength for small system\cite{GuoHM14PLA}. The quasiparticle end state indicates that the Mott state is topological. The topological property of the Mott state can be encoded by the Berry phase using the twisted boundary condition. The twisted boundary condition is defined as $\left\vert \psi_N^g\left(  j+L,\delta,\theta\right) \right\rangle =e^{i\theta}\left\vert \psi_N^g\left(  j,\delta,\theta\right) \right\rangle $ where $j$ denotes an arbitrary site, $\theta$ is the twist angle and takes values from $0$ to $2\pi$. The Berry phase is written as
\begin{equation}
\gamma={\oint}i\left\langle \psi_N^g\left(  \theta\right)  \left\vert \frac{d}{d\theta}\right\vert \psi_N^g\left(  \theta\right)  \right\rangle
\end{equation}

Our calculation demonstrates that the Berry phase $\gamma=\pi$ for system size $L=20$ using the exact diagonalization. On the other hand, in Fig.\ref{Fig1}$c$, we show the electron density ($\left\langle{n}_{j}\right\rangle=\left\langle \psi_{N}^{g}\left\vert n_{j}\right\vert\psi_{N}^{g}\right\rangle$) for the system with $L=100, V=1, \delta=0, \lambda=0.9, \nu=1/2,N=50$. Using the formula $Q_L=\sum_{j=1}^{L/2}(\langle c_{j}^{\dagger}c_{j}\rangle-\bar{\rho})$ (here, $\bar{\rho}$ is the bulk charge density), we found that the left \textit{end charge is fractional and quantized as $-e/2$}. The other end charge is fractional quantized $e/2$. These end charges are mainly localized at the two ends, as shown in Fig.\ref{Fig1}$c$.  In short, this Mott state is topological and the periodically modulated interaction (dimer interaction) can induce the \textit{metal-topological-Mott-insulator phase transition}.

\subsection{Topological phase transition and fractal-like structure}\label{PhaseButterfly}

\begin{figure}[tbhp]
\center{\includegraphics[scale=0.5]{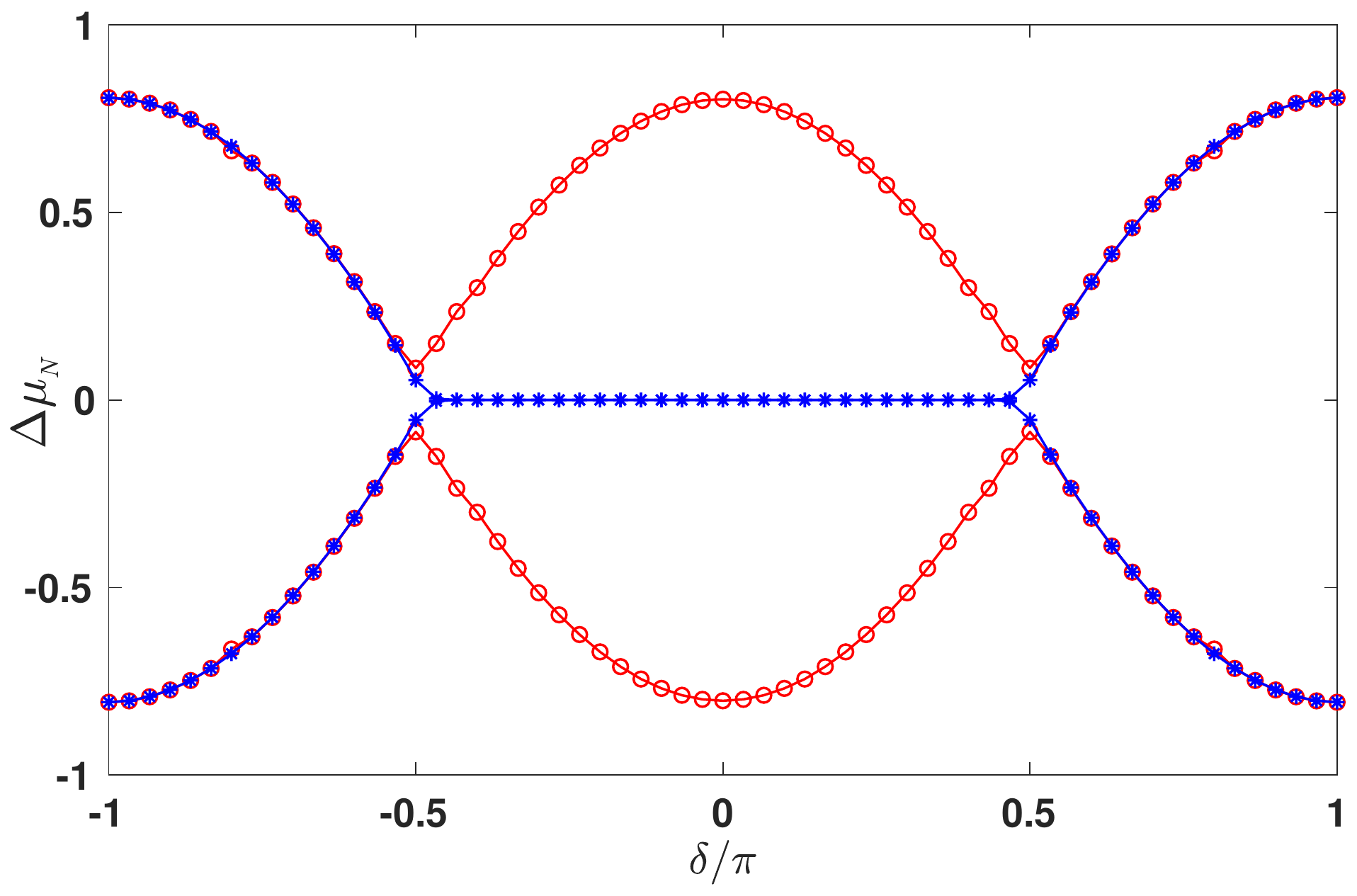}}
\caption{(Color online) The quasiparticle energy spectrum $\Delta\mu_N$ ($N=49, 50$) with respect to phase parameter $\delta$ for the system with $L=100, V=1, \alpha=1/2, \lambda=0.9$, under PBC (red) and OBC (blue).}
\label{FigDeta}
\end{figure}

To systematically study the topological properties of the system, we change the phase parameter $\delta$. Firstly, we defined the quasiparticle energy spectrum $\Delta\mu_{N}=E_{0}(L, N+1)-E_{0}(L, N)$. In Fig.\ref{FigDeta}, we show the quasiparticle energy spectrum with respect to the $\delta$ for the system with lattice $L=100$, interaction strength $V=1$,  modulation period $\alpha=1/2$,  modulation amplitude $\lambda=0.9$, under PBC (red) and OBC (blue). The quasiparticle energy spectrum split into two branches separated by a finite gap under PBC, whereas there are clearly exist two degenerate zero modes states in $( -\pi/2, \pi/2)$ region under OBC. These gapless zero modes states are closely resembling the appearance of edge states in single-particle spectra of topological Bloch bands, signaling the nontrivial topological properties of the Mott insulator. We can numerically demonstrate that these in-gap zero modes correspond to the topological end states. In Fig.\ref{Fig1}$b$, the density distribution corresponds to the zero mode at phase $\delta=0$, and particles number $N=50$ in Fig.\ref{FigDeta}. What's more, we find Berry phase $\gamma=\pi$ in these zero modes region and $\gamma=0$ for other region. In a word, the Mott (CDW) states in $(-\pi/2,\pi/2)$ region are topological nontrivial and trivial for other regions. The system can change from topological Mott state to conventional Mott state by adjusting the phase parameter $\delta$. There are two lattice sites in a unit cell when modulation period $\alpha=1/2$.  In repulsive regime, the system is in topological Mott states with Berry phase $\gamma=\pi$ when the intracellular interaction strength is bigger than the intercellular interaction strength ($V_{1}<V_{2}$) for these parameters $\nu=\alpha=1/2$. The system is in conventional Mott states for $V_{1}>V_{2}$. At the $\delta=\pm\pi/2$, the system reduces to ordinary spinless fermion interacting case and evolves into gapless state at current parameters. These behaviors are similar to the single-particle SSH model. In the topological Mott states, the fractional end charges are quantized $\pm e/2$ and mainly localized at the two ends. In the trivial case, there is no fractional end charges.

\begin{figure}[tbhp]
\center{\includegraphics[scale=0.46]{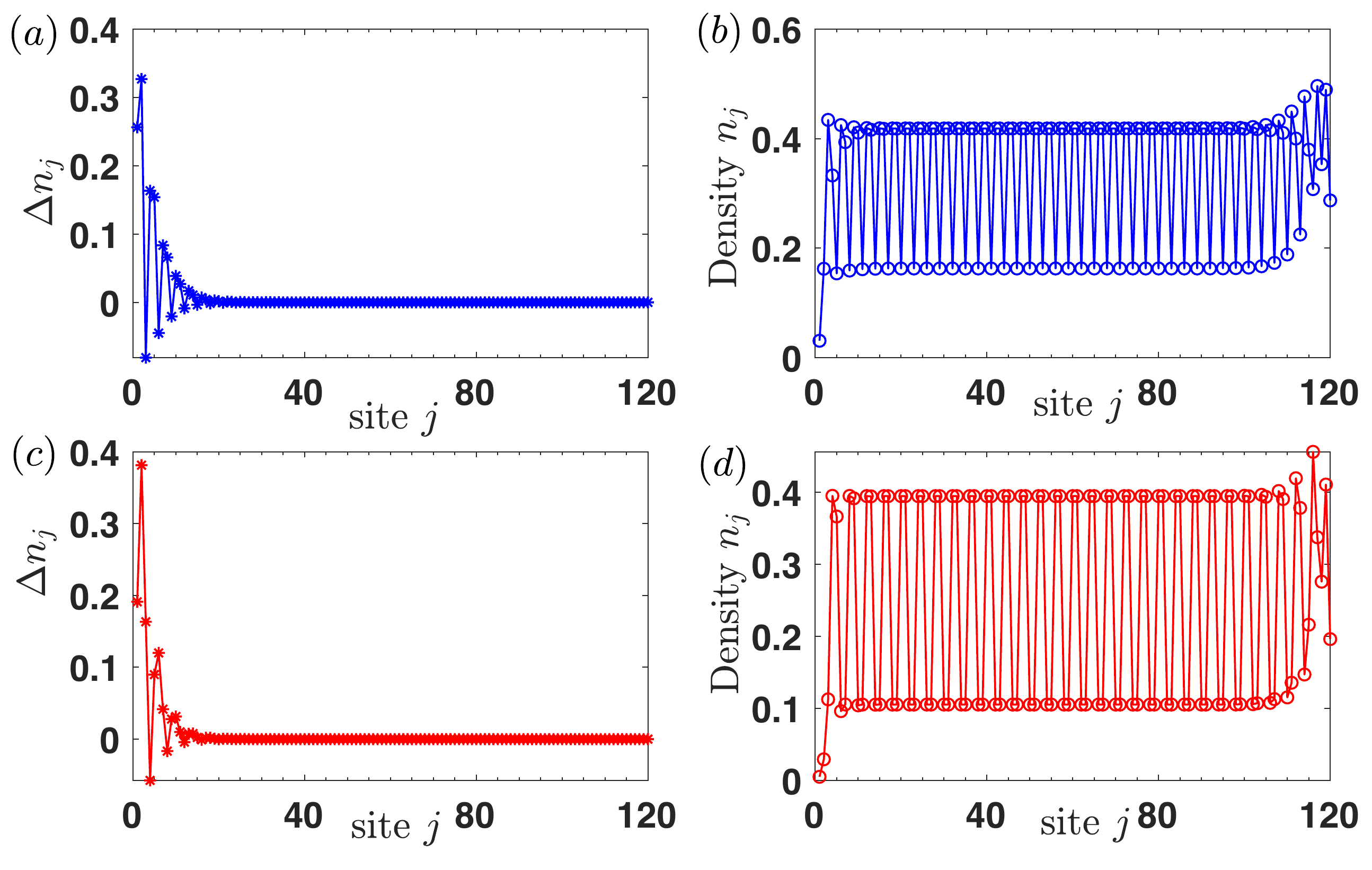}}
\caption{(Color online)The density distribution of the quasiparticle with $(a)$ $\alpha =\nu=1/3$, and $(c)$ $\alpha=\nu=1/4$.  The electron density distribution for  $(b)$ $\alpha =\nu=1/3$, and $(d)$ $\alpha=\nu=1/4$. The other parameters are $L=120, V=1, \delta=0$, and $\lambda=0.9$.}
\label{FigEnd34}
\end{figure}

The topological properties for the system as the half filling $\nu=1/2$, and modulated period $\alpha=1/2$ have been confirmed by calculating the Berry phase and quasiparticle end states. Now, we move to other periodically modulated interaction cases, considering the modulated period $\alpha=\nu=1/3$, and $1/4$ cases. In Fig.\ref{FigEnd34} (a, c), we can see clearly that the  density distribution of the quasiparticle is mainly localized at one end of chain for modulated period $\alpha=1/3$, and $1/4$. The fractional end charges are also quantized $\pm e/2$ for the two cases, as shown in Fig.\ref{FigEnd34} (b, d). The Berry phases $\gamma$ are $\pi$ for these two modulated periods calculated by exact  diagonalization. The topological properties of the Mott states at modulated period $\alpha=1/3$, and $1/4$ can also be decoded by the Chern number. The Chern number for our many-body interaction states can be defined as an integral invariant $C=\frac{1}{2\pi }\int d\delta d\theta F\left(  \delta,\theta\right)  $, where $F\left(\delta,\theta\right)  =Im\left(  \left\langle \partial_{\theta}\psi_N^g\left.  {}\right\vert \partial_{\delta}\psi_N^g\right\rangle -\left\langle \partial_{\delta}\psi_N^g\left.  {}\right\vert \partial_{\theta}\psi_N^g\right\rangle \right) $ is the Berry curvature. The Chern number $C=1$ for the system at filling factors $\nu=\alpha=1/3$ and $\nu=\alpha=1/4$. The topological properties of $\nu=2/3$ at $\alpha=1/3$ and $\nu=1/2, 3/4$ at $\alpha=1/4$ can be analyzed by similar methods.

\begin{figure}[t]
\center{\includegraphics[scale=0.65]{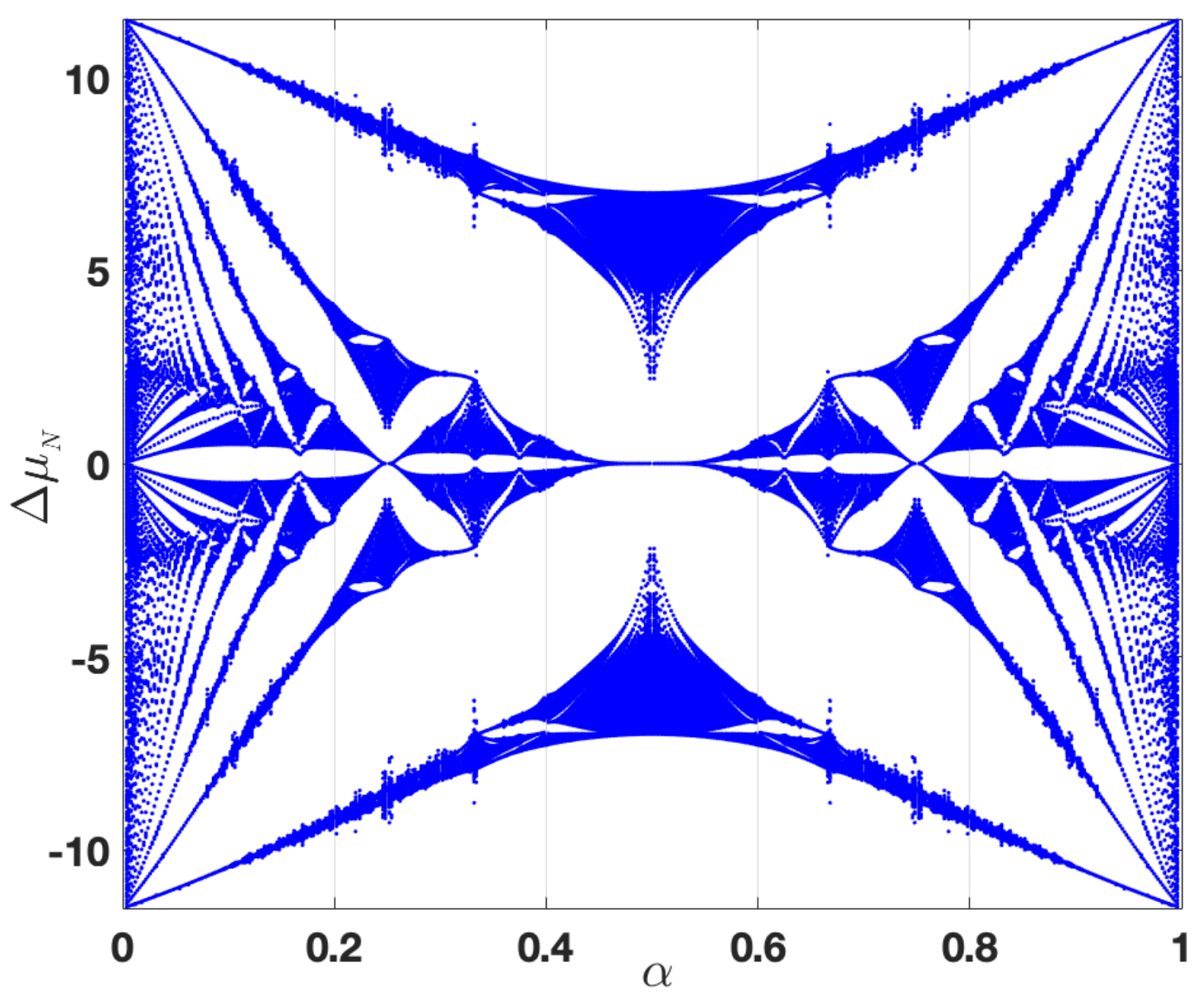}}
\caption{(Color online) The fractal-like quasiparticle energy spectra $\Delta\mu_N$ with modulated interaction parameter $\alpha$ for the system with $L=500, V=5, \delta=0$, and $\lambda=0.9$ under PBC.}
\label{FigButterfly}
\end{figure}

Next, we consider the more general modulated interaction parameter $\alpha$ cases. In Fig.\ref{FigButterfly}, we plot the quasiparticle energy spectra $\Delta\mu_N$ ($1<N<500$) as a function of periodically modulated interaction period $\alpha$ for the system with $L=500, V=5, \delta=0$, and $\lambda=0.9$ under PBC. The quasiparticle energy spectra forms a stunning fractal-like structure, which is analogous to the single-particle Hofstadter butterfly\cite{Hofstadter76PRB}. This fractal-like structure takes place in 1D spinless many-body system. The modulated parameter $\alpha$ and magnetic flux in Hofstadter model play same role on the fractal-like structure. From the fractal-like quasiparticle energy spectra, we can see whether the system is insulating state or not at various fractional fillings for a fixed modulated period. For these gap states, the Chern number (Berry phase) can be calculated and used to characterize the topological properties of the system. We can interpret the formation for the fractal-like structure in the following arguments: the interaction becomes inhomogeneous and oscillate because of the modulated parameter.  As one tunes the filling factor and Fermi momentum $k_F$ changes, the modulated interaction process can induce the energy gap at fractional filling and metal-insulator transition takes place when the Fermi momenta and the specific Fourier wave-vector components of modulated interaction commensurate.

\subsection{Quasi-periodic case}\label{Quasiperiod}

\begin{figure}[t]
\center{\includegraphics[scale=0.61]{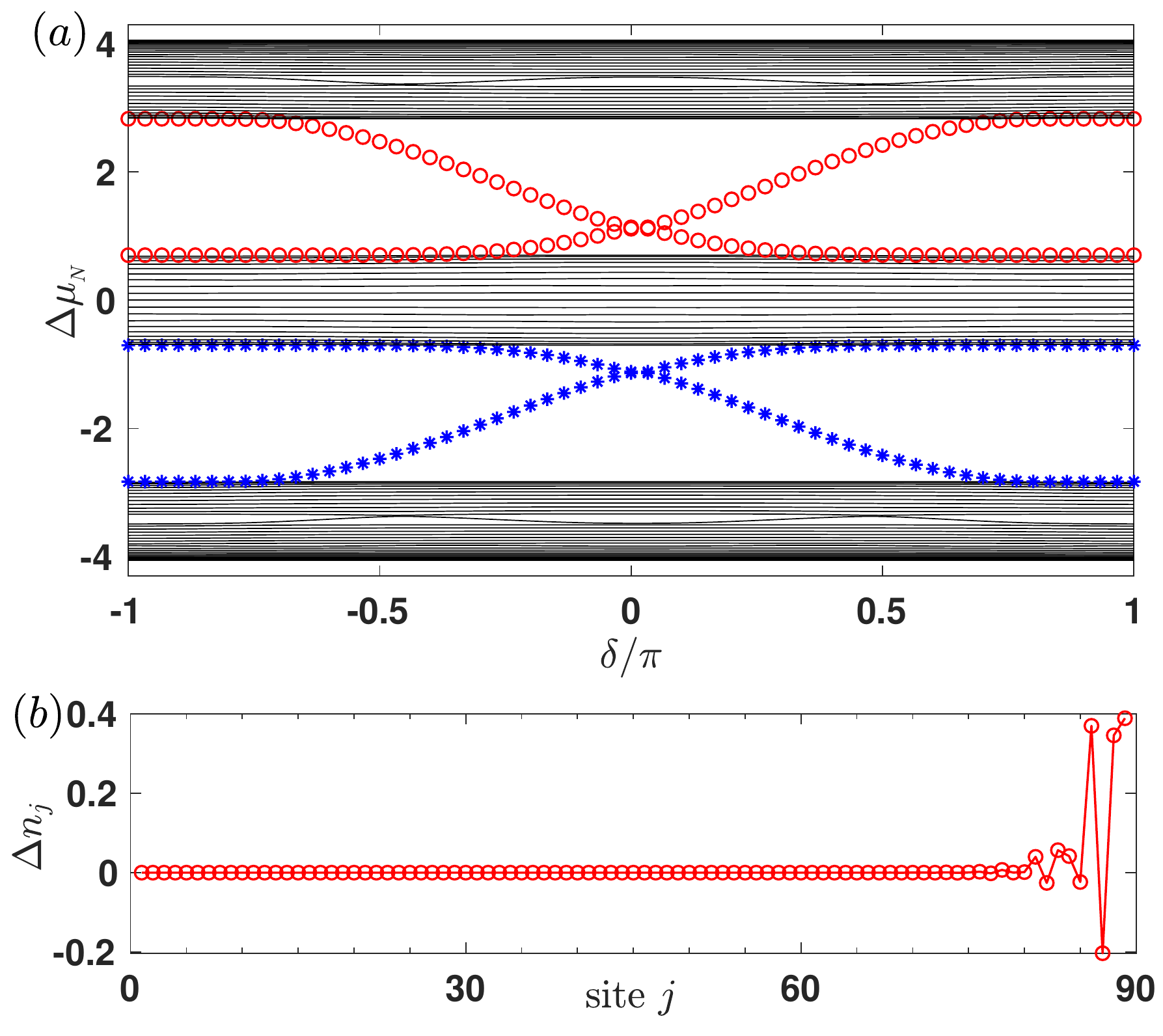}}
\caption{(Color online) $(a)$ The quasiparticle energy spectra $\Delta\mu_N$ with respect to phase parameter $\delta$ for the system under OBC; $(b)$ The density distribution of the quasiparticle with $\delta=0$. The other parameters are $L=89, N=55, V=2,\alpha=(\sqrt5-1)/2, \lambda=0.9$.}
\label{FigQuasiPeriod}
\end{figure}

Encouraged by the success of periodically modulated interaction case, we now turn to the quasi-periodic case. Without loss of generality, we choose the $\alpha=(\sqrt{5}-1)/2$ as an illuminating example. In Fig.\ref{FigQuasiPeriod} $(a)$, the quasiparticle energy spectra $\Delta\mu_N$ is shown for the system $L=89$ about the phase parameter $\delta$ under OBC. There are crossing gapless in-gap states at the specific filling factors $\nu\approx\alpha$ and $\nu \approx1-\alpha$, which are the gapped CDW states under PBC. Taking the system with $L=89,N=55$ as an example, we find that they are the localized end quasiparticle states, as shown by Fig.\ref{FigQuasiPeriod} $(b)$, The topological properties of the CDW states can be further confirmed by computing the Chern number. The Chern number of the system with $L=21$ are $1$ at $\nu\approx\alpha$ ($N=13$) filling and $-1$ at $\nu\approx1-\alpha$ ($N=8$) filling. So, the quasi-periodic modulated interaction can also induce the system into the incommensurate topological Mott (CDW) states.

\section{Conclusion}\label{Conclusion}

In a word, we demonstrate that the spatially modulated interaction can be a new elegant avenue towards realizing the interacting topological phases. We take a paradigmatic model consisting of a one-dimensional spinless fermion lattice with spatially modulated interaction as an example. We found that the periodically modulated interaction can induce the topological-Mott-insulator phase transition. The topological Mott state is coded by quasiparticle end states and the fractional quantized $e/2$ end charges, and characterized by Berry phase and Chern number. The quasiparticle energy spectra as a function of modulated interaction period appears a stunning fractal-like structure. The topological Mott state can also appear in the quasi-period modulated interaction case. There are a number of future directions emanating from our work. Recently, there has been significant interest in 1D spinless fermion system with attractive interactions\cite{RuhmanJ15PRL,Kane17PRX,VyborovaVV18PRB, HeYC19PRB}. It is therefore of substantial interest and importance that investigating the attractive regime for the spatially modulated interaction. Another natural question concerns the effect of disorder. According to the Jordan-Wigner transformation, the modulated interaction spinless fermion model is equivalent a $S=1/2$ anisotropy-modulated XXZ spin chain model. The anisotropy-modulated XXZ spin chain is related to the quantum spin chain with periodic modulated exchange coupling\cite{HuHP14PRB,HuHP16PRB,LadoJL19PRR}. On the other hand, our model can also be mapped to the hard-core bosons model, via Holstein-Primakoff transformation for spin-1/2 particles. Thus, the modulated interaction spinless fermions model, and the corresponding spin chain and hard-core bosonic models should share the same physics. Finally, the modulated interaction procedure presented here could be generalized to other systems, although the details may be more complicated.

\ack
We thank J.K. Jain, and Chao-Xing Liu for helpful discussions. This work was supported by the National Natural Science Foundation of China (NSFC) under grant numbers 11604081, 11447008. Z.W.Z. is grateful to the China Scholarship Council for financial support.

\section*{References}

%\bibliographystyle{iopart-num}
%\bibliography{TopologicalInsulators}
\providecommand{\newblock}{}

\end{document}